\documentclass[12pt]{article}

\usepackage[english]{babel}
\usepackage{amsfonts}
\usepackage{amssymb}
\usepackage{graphicx}

\newcommand{\be}{\begin{equation}}
\newcommand{\ee}{\end{equation}}
\newcommand{\bea}{\begin{eqnarray}}
\newcommand{\eea}{\end{eqnarray}}

\begin{document}

\begin{titlepage}

\begin{flushright}
{\tt
    hep-th/0512181}
 \end{flushright}

\bigskip

\begin{center}

{\Large \bf{ Regularity of the stress-energy tensor for extremal
Reissner-Nordstr\"om black holes}}

\bigskip
\bigskip\bigskip
S. Farese\footnote{farese@ific.uv.es}\footnote{Talk given at the
conference ``Constrained dynamics and quantum gravity", Cala Gonone (Italy), 
September 2005}

\end{center}

\bigskip%

\footnotesize \noindent {\it  Departamento de F\'{\i}sica
Te\'orica and
    IFIC, Centro Mixto Universidad de Valencia-CSIC.
    Facultad de F\'{\i}sica, Universidad de Valencia,
        Burjassot-46100, Valencia, Spain.}

\bigskip

\bigskip
\bigskip \bigskip \bigskip 
\begin{center}
{\bf Abstract}
\end{center}
We calculate the expectation values of the stress-energy tensor for
both a massless minimally-coupled and dilaton-coupled 2D field
propagating on an extremal Reissner-Nordstr\"om black hole, showing
its regularity on the horizon in contrast with previous claims in
the literature.

\bigskip


\end{titlepage}

\newpage

\section{Introduction}
Hawking discovery of particles production by black holes
\cite{Hawking}, gave rise to a great investigation of quantum
effects in strong gravitational fields.\\
As we do not possess a full quantum theory of gravitation, the
framework in which we work is the so-called semiclassical
approximation, in which the gravitational field is treated at the
classical level, whereas the matter fields are quantized. The
classical energy-momentum tensor in the r.h.s of Einstein's
equations is replaced by its quantum expectation values in a
suitable quantum state: $G_{\mu\nu}(g_{\alpha\beta})=8\pi\langle
T_{\mu\nu}(g_{\alpha\beta})\rangle$ (in units $\hbar=c=G=1$). The
main problem is just to find a covariant expression for the quantum
stress-energy tensor in a general curved spacetime. The exact form
of $\langle T_{\mu\nu}(g_{\alpha\beta})\rangle$ is in general
unknown in four dimensions and therefore, with the hypothesis of
spherical symmetry in 4D for the metric and selecting the $s$-wave
component of the matter field, we work in a
two-dimensional spacetime.\\
We study the energy-momentum tensor of a quantum scalar field
propagating on an extremal Reissner-Nordstr\"om black hole. The
interest in such black holes finds its justification in the fact
that, being characterized by zero Hawking temperature, they are
supposed not to emit particles and so they can be considered as
the final state of the evaporation of nonextremal ones.\\
In the literature it is shown that, in two dimensions, the $\langle
T_{\mu\nu}(g_{\alpha\beta})\rangle$ of a scalar field is divergent
on the horizon of extremal black holes \cite{Trivedi,Loranz}. Such
divergence is physically unacceptable because, as we said, the
quantum stress-energy tensor is supposed to be the source term of
the semiclassical Einstein's equations. The question we shall try to
answer now is if this divergence is indeed realized in a physically
realistic situation.

\section{Scalar field in \emph{s}-wave approximation}
Let us consider a four dimensional massless minimally-coupled scalar
field, which has the following classical action \be \label{4daction}
S^{(4)}=-\frac{1}{8\pi}\int d^4x\sqrt{-g^{(4)}}(\nabla f)^2. \ee A
spherically symmetric four-dimensional metric can be written as \be
\label{sphsymetric} ds_{(4)}^2=ds_{(2)}^2+r_0^2 e^{-2\phi}d\Omega^2,
\ee where we have parameterized the radius of the two-sphere
$r^2=r_0^2 e^{-2\phi}$ through a dilaton field $\phi$ and $r_0$ is
an arbitrary scale factor. In a spherically symmetric background any
field can be expanded in spherical harmonics, of which we pick up
only the \emph{s}-wave component: $f=f(t,r)$. With these hypothesis,
the 4D action can be integrated with respect to the angular
coordinates, obtaining so a two-dimensional action \be
\label{2daction} S^{(2)}=-\frac{1}{2}\int
d^2x\sqrt{-g^{(2)}}e^{-2\phi}(\nabla f)^2, \ee where $g^{(2)}$
represents the radial sector of the metric. We remark that the
scalar field acquires now a non trivial coupling with the dilaton
field besides the usual one to the metric.

\section{Near horizon approximation} As we are interested in the
behavior of the stress-energy tensor on the horizon of a black hole,
we first consider the near horizon approximation. Fixing the radius
of the two-sphere equal to the horizon's one, $r_0=r_H, \phi=0$, we
obtain just the action of a 2D scalar field \be
S^{(2)}=-\frac{1}{2}\int d^2x\sqrt{-g^{(2)}}(\nabla f)^2. \ee It is
useful to work in conformal coordinates:
$ds_{(2)}^2=-e^{2\rho(x)}dx^+ dx^-$. The classical action satisfies
two invariance laws: general covariance, to which it corresponds a
conserved stress-energy tensor ($\nabla_\mu T^{\mu\nu}=0$), and
Weyl-invariance, which has as
consequence a traceless energy-momentum tensor.\\
An intuitive way to derive the quantum stress-energy tensor is by
the use of Virasoro anomaly and the equivalence principle (for
details see \cite{Fabbri_Navarro-Salas}). One obtains the following
expression for the expectation value of the stress tensor in a
generic state $|\Psi\rangle$ \be\label{pst}
\langle\Psi|T_{\pm\pm}(x^\pm)|\Psi\rangle
=\langle\Psi|:T_{\pm\pm}(x^\pm):|\Psi\rangle-\frac{1}{12\pi}(\partial_{\pm}\rho\partial_{\pm}\rho
-\partial_{\pm}^2\rho).\ee At the quantum level it is impossible to
preserve the two invariances. Therefore, preserving general
covariance and imposing the stress-energy tensor conservation, the
Weyl-invariance is broken and an anomalous trace appears
\be\label{trace} \langle T \rangle=\frac{1}{24\pi}R. \ee
Alternatively, one can obtain the quantum stress tensor by variation
of the non local Polyakov action: $ S_P=-1/(96\pi)\int
d^2x\sqrt{-g^{(2)}}R\Box ^{-1} R$.\\
From the expression of the quantum stress-energy tensor (Eq.
(\ref{pst})), one can obtain the relation between its expectation
values in two different vacuum states $|x^\pm\rangle$ and
$|\tilde{x}^\pm\rangle$: \be \label{pvc} \langle
\tilde{x}^\pm|T_{\pm\pm}(x^\pm)|\tilde{x}^\pm\rangle= \langle
x^\pm|T_{\pm\pm}(x^\pm)|x^\pm\rangle-\frac{1}{24\pi} \{
\tilde{x}^{\pm}, x^{\pm} \},\ee where $|x^\pm\rangle$
($|\tilde{x}^\pm\rangle$) represents the vacuum state determined by
the expansion of the field in the modes which  are positive
frequency  with respect to the time $(x^++x^-)/2$
($(\tilde{x}^++\tilde{x}^-)/2$) and $\{ \tilde{x}^{\pm}, x^{\pm}
\}=\frac{d^3\tilde{x}^\pm}{dx^{\pm3}}\big/\frac{d\tilde{x}^\pm}{dx^\pm}
-\frac{3}{2}\bigg(\frac{d^2\tilde{x}^\pm}{dx^{\pm2}}/\frac{d\tilde{x}^\pm}{dx^\pm}\bigg)^2$
is the Schwarzian derivative associated to the conformal
transformation $\{x^\pm\}\to\{\tilde{x}^\pm\}$.

\section{Nonextremal Reissner-Nordstr\"om black\\ holes}
We shall now apply the formalism developed above to the case of the
propagation of a scalar field in the background of a charged black
hole, which is characterized by the following line element \be
ds^2=-f(r)dudv=-\bigg(1-\frac{2M}{r}+\frac{Q^2}{r^2}\bigg)dudv \ee
and possesses two horizons: $r_\pm=M\pm\sqrt{M^2-Q^2}$ (with $M$ and
$Q$ respectively the mass and the charge of the black hole). We
shall analyze the behavior of the energy-momentum tensor on the
outer
horizon $r_+$.\\
In the study of black holes, one usually consider three vacuum
states.
\begin{itemize}
\item \emph{Boulware state}
\end{itemize}
It is constructed by expanding the field in modes which are positive
frequencies with respect to the asymptotically Minkowskian time
coordinate: $(4\pi\omega)^{-1/2}e^{-i\omega
v},\quad(4\pi\omega)^{-1/2}e^{-i\omega u}$. These modes reduce at
infinity to plane waves, so the Boulware state reproduces, in the
limit $r\to\infty$, the notion of vacuum in flat spacetime. This
state is supposed to describe the vacuum polarization outside
a static star, whose radius is bigger than the horizon's one.\\
With the use of Eq. (\ref{pst}), with $\rho=\frac{1}{2}\ln f$, we
can evaluate the energy-momentum tensor in this state \be \langle
B|T_{uu}|B\rangle=\langle B|T_{vv}|B\rangle=\frac{1}{24\pi
r^3}\bigg(-M+\frac{3}{2}\frac{M^2+Q^2}{r}-\frac{3MQ^2}{r^2}+\frac{Q^4}{r^3}\bigg).\ee
As it is plausible to expect, the stress-energy tensor in this sates
goes to zero at infinity. To check the regularity on the horizon ,
we have to express its components in a coordinate system which is
regular there. To this end, we introduces the Kruskal coordinates
\be \label{kruskal} U=-\frac{1}{\kappa_+}e^{-\kappa_+ u}\quad,\quad
V=\frac{1}{\kappa_+}e^{\kappa_+ v},\ee where
$\kappa_+=\frac{1}{2}\frac{\partial f}{\partial
r}|_{r_+}=\sqrt{M^2-Q^2}/r_+^2$ is the surface gravity on the outer
horizon. With these coordinates, the regularity conditions on the
future horizon become \cite{Christensen} \be\label{regcond}
f^{-2}\langle T_{uu}\rangle <\infty\quad,\quad\langle
T_{vv}\rangle<\infty\quad,\quad f^{-}\langle T_{uv}\rangle
<\infty\ee and the same with just $u$ and $v$ interchanged on the
past horizon. It is easy to check that the Boulware state is
divergent on both the future and the past horizons.
\begin{itemize}
\item \emph{Hartle-Hawking state}
\end{itemize}
This state is constructed by expanding the field in ingoing
(outgoing) modes which are positive frequencies with respect to the
affine parameter on the future (past) horizon:
$(4\pi\omega)^{-1/2}e^{-i\omega
V},\quad(4\pi\omega)^{-1/2}e^{-i\omega U}$. One can evaluate the
quantum stress-energy tensor in this state, starting by the
corresponding one in the Boulware state, by means of relation
(\ref{pvc}) \be \langle H|T_{uu}|H\rangle=\langle
B|T_{uu}|B\rangle-\frac{1}{24\pi} \{U,u\}=\langle
B|T_{uu}|B\rangle+\frac{1}{48\pi}\kappa_+ ^2=\langle
H|T_{vv}|H\rangle.\ee The difference between the two vacuum states
is induced by the Schwarzian derivative, which, by virtue of the
exponential relation between Eddington-Finkelstein and Kruskal
coordinates, turns out to be proportional to the square of the
surface gravity. The Hartle-Hawking state satisfies the regularity
conditions (\ref{regcond}), but at infinity it is different from
zero: the asymptotic limit represents thermal radiation at the
Hawking temperature $T_H$ \be \langle H|T_{uu}|H\rangle=\langle
H|T_{vv}|H\rangle\stackrel{I^+}{\longrightarrow}\frac{1}{48\pi}\kappa_+
^2=\frac{\pi}{12}T_H^2.\ee The physical interpretation of this state
is that of thermal equilibrium of the black hole with its own
radiation.
\begin{itemize}
\item \emph{Unruh state}
\end{itemize}
It is constructed by choosing ingoing modes positive frequency with
respect to the asymptotic Minkowski time, and outgoing modes
positive frequency with respect to the affine parameter on the past
horizon: $(4\pi\omega)^{-1/2}e^{-i\omega v}$,
$(4\pi\omega)^{-1/2}e^{-i\omega U}$. This state represents the
vacuum of particles incoming from $I^-$. Incoming modes propagating
on the geometry of a collapsing body, after being reflected at the
origin, emerge, just before the horizon formation, exponentially
redshifted as positive frequency with respect to $U$. This "hybrid"
construction describes the late-time behavior of a quantum field in
the spacetime of a collapsing body forming a black hole. The
stress-energy tensor turns out to be \bea \langle
U|T_{uu}|U\rangle&=&\langle H|T_{uu}|H\rangle=\langle
B|T_{uu}|B\rangle+\frac{1}{48\pi}\kappa_+ ^2,\\
\langle U|T_{vv}|U\rangle&=&\langle B|T_{vv}|B\rangle.\eea In the
case of a realistic collapse one should consider a different vacuum
state, the so-called \emph{in} vacuum, containing extra terms
depending on the details of the collapse, which drop exponentially
to zero at late time, leaving just the constant contribution of the
Schwarzian derivative: \be \langle
in|T_{\mu\nu}|in\rangle\stackrel{r\to r_+}{\longrightarrow}\langle
U|T_{\mu\nu}|U\rangle+O(e^{-\kappa_+ u}).\ee The Unruh state is
regular on the future horizon. At future infinity it gives a
constant outgoing flux which represents the asymptotical value of
Hawking radiation \be \langle
U|T_{vv}|U\rangle\stackrel{I^+}{\longrightarrow}\frac{1}{48\pi}\kappa_+
^2.\ee

\section{Extremal Reissner-Nordstr\"om black holes}
Let us consider now the more peculiar case of the extreme charged
black holes: \be ds^2=-f(r)dudv=-\bigg(1-\frac{M}{r}\bigg)^2dudv.\ee
The two horizons now coincide at the value $r_H=M$. The surface
gravity is null and, being zero the Schwarzian derivatives $\{U,u\}$
and $\{V,v\}$, it could seem that the three vacuum states all
coincide \be \langle\label{hextr} T_{uu}\rangle=\langle
T_{vv}\rangle=-\frac{1}{24\pi}\frac{M}{r^3}\bigg(1-\frac{M}{r}\bigg)^3\equiv
H^{extr}(r).\ee This stress-energy tensor does not satisfy the
regularity conditions and we have the problem of the divergence on
the horizon. However, we have to note that the Kruskal coordinates
we have used to construct a regular state on the horizon (see
(\ref{kruskal})), are not defined in the limit $\kappa_+\to 0$. To
analyze the problem, we have to understand if the quantum state
whose energy-momentum tensor in given by Eq. (\ref{hextr}) is
realizable by means of a physical process. To this end, let us
consider the process of formation of an extremal black hole,
starting by a flat spacetime,  by means of the collapse of an
ingoing charged null shell located at $v=v_0$ (see.
Fig.1)\cite{Balbinot...}.
\begin{figure}
\begin{center}
\includegraphics[angle=0,width=1.5in,clip]{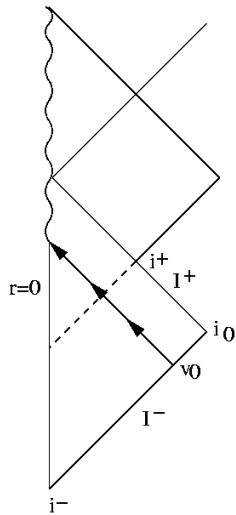}
\caption{Penrose diagram describing the formation of an extreme
Reissner-Nordstr\"om black hole.} \label{fig1}
\end{center}
\end{figure}
The metric will be the following \bea
v<v_0\quad&\quad&ds^2=-du_{in}dv_{in}\quad\quad\qquad \mathrm{Minkowski\ spacetime},\nonumber\\
v>v_0\quad&\quad&ds^2=-\bigg(1-\frac{M}{r}\bigg)^2dudv\quad\mathrm{Extremal\
black\ hole}.\nonumber\eea Matching the two metrics across the
shell, we obtain the relation $u(u_{in})$ \be u=u_{in}-4M
\bigg[\ln\big(\frac{v_0-u_{in}}{2M}-1\big)-\frac{1}{2\big(\frac{v_0-u_{in}}{2M}-1\big)}\bigg].
\ee A positive frequency $v_{in}$ mode at $I^-$, after being
reflected at $r=0$, becomes at late retarded time $(u\to+\infty)$ a
positive frequency $U$ mode, where \cite{Liberati} \be u=-4M
\bigg[\ln\bigg(-\frac{U}{M}\bigg)+\frac{M}{2U}\bigg]\qquad\mathrm{(with}\
u_{in}\to 2U-2M+v_0).\ee With this new Kruskal-type coordinate, the
\emph{in} vacuum state turns out to be \bea \langle
in|T_{uu}|in\rangle&=&H^{extr}(r)-\frac{1}{24\pi}
\{U,u\}\nonumber\\
&=&-\frac{1}{24\pi}\frac{M}{r^3}\bigg(1-\frac{M}{r}\bigg)^3+\frac{1}{24\pi}\frac{U^3(U-2M)}{2M^2(2U-M)^4},\\
 \langle
in|T_{vv}|in\rangle&=&H^{extr}(r).\eea The stress-energy tensor is
now regular on the future horizon $(U\sim-(r-M))$: \bea\lim_{r\to
M}\bigg(\frac{du}{dU}\bigg)^2\langle
in|T_{uu}|in\rangle&=&-\lim_{r\to
M}\frac{1}{24\pi}\bigg[\frac{M}{r^2}\frac{1}{r-M}-\frac{1}{M(r-M)}+finite
\bigg]\nonumber\\
&=&-\frac{1}{24\pi}\frac{3}{2M^2}<\infty.\eea The divergence coming
from the vacuum polarization part is canceled by the divergent term
induced by the Schwarzian derivative. Let us note that in this case
the transient terms in the stress-energy tensor for the \emph{in}
vacuum do not drop to zero exponentially, but with power law;
however, unlike the nonextremal case, one can not simply discard
this contribution to get the late-time behavior, since this
procedure would lead to the incorrect result we have previously
seen. Also the vacuum polarization vanishes at late retarded time
and it is necessary to
consider both terms to have regularity on the future horizon.\\
The result has also been generalized to the case of the collapse of
a timelike shell \cite{Fagnocchi-Farese}. In this case the relation
$u(U)$ is \be u=-4M \bigg[\ln\bigg(-\frac{U}{c M}\bigg)+\frac{c
M}{2U}\bigg],\ee where $c$ is a constant related to the details of
the collapse which does not affect the regularity of the
stress-energy tensor on the horizon.

\section{Dilatonic theory}
The result obtained in the previous section can be extended to the
most general case of a dilaton-coupled scalar field, whose
classical action is given by (\ref{2daction}).\\
Exactly as we did for the Polyakov theory, we can derive the
expression for the quantum expectation values of the energy-momentum
tensor by means of the generalized Virasoro anomaly and the
equivalence principle \cite{Fabbri-Farese-NavarroSalas} \bea
\label{qstdilaton3} \langle\Psi|T_{\pm\pm}(x^+,x^-)|\Psi\rangle &=&
\langle\Psi|:T_{\pm\pm}(x^+,x^-):|\Psi\rangle\\
&&-\frac{1}{12\pi}(\partial_{\pm}\rho\partial_{\pm}\rho
-\partial_{\pm}^2\rho)+ \frac{1}{2\pi} \left[  \partial_{\pm}\rho
\partial_{\pm}\phi + \rho (\partial_{\pm}\phi)^2 \right].\nonumber
\eea We note that new local terms depending on the dilaton field
appear.\\
From the expression of the quantum stress-energy tensor we obtain
the following relation between the expectation values of the same
tensor in two different vacuum states
 \bea \label{transf2}
\langle \tilde{x}^\pm|T_{\pm\pm}(x^+,x^-)|\tilde{x}^\pm\rangle&=&
\langle x^\pm|T_{\pm\pm}(x^+,x^-)|x^\pm\rangle-\frac{1}{24\pi}
\{ \tilde{x}^{\pm}, x^{\pm} \}\\
&&+\frac{1}{4\pi}
\ln\bigg(\frac{dx^-}{d\tilde{x}^-}\frac{dx^+}{d\tilde{x}^+}\bigg)(\partial_\pm
\phi
)^2\nonumber\\
&&+\frac{1}{4\pi}\frac{d^2x^\pm}{d\tilde{x}^{\pm2}}\bigg(\frac{dx^\pm}{d\tilde{x}^\pm}\bigg)^{-2}
\partial_{\pm}\phi.\nonumber\eea
Evaluating the $\langle in|T_{UU}|in\rangle$ component of the
stress-energy tensor for the extremal Reissner-Nordstr\"om geometry,
we can see that new divergent terms, both polynomial and
logarithmic, appear in addition to those already present in the
Polyakov case. However it is easy to show (see
\cite{Fagnocchi-Farese} for more details) that the polynomial terms
exactly cancel each other on the horizon and the logarithmic ones
too sum up to give a finite value. Therefore the stress-energy
tensor turn out to be finite on the future horizon in this more
involved case too.

\section{Conclusions}
Unlike previous claims in the literature, we have shown the
regularity of the energy-momentum tensor at the horizon of an
extremal Reissner-Nordstr\"om black hole formed by gravitational
collapse in the two cases of a massless minimally-coupled 2D scalar
field, and a massless dilaton-coupled 2D scalar field. Besides we
have shown that the regularity is independent on the details of the
collapse.

\end{document}